\documentclass[12pt]{article}
\usepackage{a4wide}
\usepackage{amssymb}
\usepackage{graphicx}
\begin{document}
{\renewcommand{\thefootnote}{\fnsymbol{footnote}}
\begin{center}
{\LARGE  A no-go result for covariance\\[2mm] in models of loop quantum gravity}\\
\vspace{1.5em}
Martin Bojowald\footnote{e-mail address: {\tt bojowald@gravity.psu.edu}}
\\
\vspace{0.5em}
Institute for Gravitation and the Cosmos,\\
The Pennsylvania State
University,\\
104 Davey Lab, University Park, PA 16802, USA\\
\vspace{1.5em}
\end{center}
}

\setcounter{footnote}{0}

\begin{abstract}
  Based on the observation that the exterior space-times of Schwarzschild-type
  solutions allow two symmetric slicings, a static spherically symmetric one
  and a timelike homogeneous one, modifications of gravitational dynamics
  suggested by symmetry-reduced models of quantum cosmology can be used to
  derive corresponding modified spherically symmetric equations. Generally
  covariant theories are much more restricted in spherical symmetry compared
  with homogeneous slicings, given by $1+1$-dimensional dilaton models if they
  are local. As shown here, modifications used in loop quantum cosmology do
  not have a corresponding covariant spherically symmetric theory. Models of
  loop quantum cosmology therefore violate general covariance in the form of
  slicing independence. Only a generalized form of covariance with a
  non-Riemannian geometry could consistently describe space-time in models of
  loop quantum gravity.
\end{abstract}

\section{Introduction}

Models of black holes in quantum gravity are valuable not only because their
strong-field effects draw considerable physical interest, but also because
they are understood as a consequence of non-trivial dynamical properties of
space-time. Given the incomplete status of all approaches to quantum gravity,
the latter connection would, at present, seem even more important than the
former. In this sense, black-hole models in quantum gravity have a clear
advantage over models of quantum cosmology because basic cosmological
solutions work with simpler space-times characterized by exact or perturbative
spatial homogeneity with a preferred background time direction.

The connection between black holes and space-time structure is particularly
relevant in canonical, background-independent approaches, such as loop quantum
gravity. In such approaches, the structure of space-time is a derived concept
and not presupposed. Explorations in a physically motivated context can
therefore provide important insights into the viability of any specific
proposal. Even if implied quantum-gravity effects in black holes may not be
realistically observable, studying them in detail can often strengthen an
analysis of purely mathematical consistency conditions on the theory.

An important step in this direction had recently been undertaken in
\cite{Transfig}. Although the initial analysis was quickly found to be invalid
--- owing to an incorrect treatment of phase-space dependent quantum
corrections \cite{DiracPoly,ExtendedPoly,bvPoly,MassPoly,QuasiPoly}, a failure
to recognize subtleties in the asymptotic structure \cite{TransCommAs}, and
unacceptable long-term effects in astrophysically relevant solutions
\cite{LoopISCO} --- it was based on an interesting suggestion that leads to a
new and independent test of space-time structure in models of loop quantum
gravity \cite{TransComm}. Here, we elaborate on this application and use it to
demonstrate a no-go result that implies the non-covariance of any model of
loop quantum gravity, if covariance is understood in the classical way related
to slicing independence in Riemannian geometry. 

\section{Symmetries in Schwarzschild space-time}

The construction utilized in \cite{Transfig} is an application of
minisuperspace results, originally derived for models of quantum cosmology, to
a black-hole context. The most common example of this form is based on the
well-known fact that the Schwarzschild solution,
\begin{equation} \label{Schwarzschild}
 {\rm d}s^2= -\left(1-\frac{2m}{r}\right){\rm d}t^2+ \frac{{\rm d}r^2}{1-2m/r}
 + r^2\left({\rm d}\vartheta^2+\sin^2\vartheta{\rm
     d}\varphi^2\right)\,,
\end{equation}
has a homogeneous spatial slicing in the interior, where $r<R_{\rm S}=2m$ is
less than the Schwarzschild radius. In this region, $r$ can serve as a time
coordinate because the restricted line element ${\rm d}s^2|_{r={\rm const}}>0$
is positive between any two distinct points at the same value of $r$. The
$r$-dependence of the coefficients in (\ref{Schwarzschild}) therefore implies
time dependence in this region, but not spatial inhomogeneity. The resulting
homogeneous dynamics is described by the Kantowski--Sachs model \cite{KS}.

A quantum scenario of the Schwarzschild interior can therefore be constructed
by importing quantum effects found in anisotropic minisuperspace models from
quantum cosmology. Of major interest to \cite{Transfig} was the possibility
that a bounce in cosmological models, as sometimes claimed in models of loop
quantum cosmology, might then be reinterpreted as a non-singular transition
through the black-hole interior. Note, however, that most bounce claims in
loop quantum cosmology are based on oversimplified models that do not capture
the correct physics near a spacelike singularity
\cite{Infrared,NonBouncing,Claims}. Moreover, such models are often in
violation of general covariance \cite{NonCovDressed}, a conclusion that will
be strengthened by our derivations below. Specific predictions made in this
context, in particular of quantitative nature as in the example of the ratio
of masses before and after the bounce, therefore cannot be considered
reliable.

Nevertheless, it is justified to assume the modified dynamics implied by a
quantum version of the Kantowski--Sachs model as a possible substitute of the
dynamics of general relativity in the Schwarzschild interior, and then to
evaluate potential implications on qualitative features of the resulting
model. An open question even in such less ambitious studies has been how to
connect the Schwarzschild interior to a possible inhomogeneous exterior
geometry. Such a connection has become possible by the useful suggestion of
\cite{Transfig} to consider homogeneous {\em timelike} slicings in the
exterior, given by constant $r$ in (\ref{Schwarzschild}) even if $r>R_{\rm
  S}$, and apply modifications proposed in minisuperspace models. Using this
method, the authors of \cite{Transfig} constructed a modified line element
that could possibly describe the exterior geometry of a quantum-modified,
non-singular Schwarzschild black hole (or its Kruskal extension).

In order to do so, \cite{Transfig} {\em assumed} that the modified exterior is
subject to the same space-time structure as the classical theory, given by
Riemannian geometry and described by a line element such as
(\ref{Schwarzschild}) but with a modified $r$-dependence in its
coefficients. However, in background-independent models of quantum gravity, it
is not guaranteed that the structure of space-time as seen in the classical
theory remains intact. Space-time structure should rather be derived from the
theory, which would then show whether a line element of the form ${\rm
  d}s^2=g_{ab}{\rm d}x^a{\rm d}x^b$, restricted to spherical symmetry in the
present context, can indeed describe the modified quantum dynamics. In
canonical approaches to quantum gravity, such as loop quantum gravity used in
\cite{Transfig}, the task is to show that gauge transformations acting on
components of $g_{ab}$, generated by modified constraints that generate the
modified dynamics used to obtain any kind of non-singular homogeneous model,
are consistent with standard coordinate transformations of ${\rm d}x^a$
assumed in the definition of a line element. We will show that this is not the
case for modifications suggested by loop quantum cosmology.

\section{Dynamical models}

For our demonstration, we will need the relevant equations that describe the
classical and modified dynamics of the slicings involved in the construction
of \cite{Transfig}. We present these equations and our new derivations in a
form based on metric variables, which are more common and therefore more
easily accessible than the triad variables used in models of loop quantum
gravity. Our general result does not depend on the choice of variables because
it is invariant under canonical transformations. For an explicit derivation in
triad variables, see \cite{TransComm}.

\subsection{Spherical symmetry and interior geometry}

We begin with the generic form of line elements subject to the symmetries of
Kantowski--Sachs models:
\begin{equation} \label{dsKS}
 {\rm d}s^2= -N(t)^2{\rm d}t^2+ a(t)^2{\rm d}x^2+ b(t)^2
 \left({\rm 
     d}\vartheta^2+ \sin^2\vartheta{\rm d}\varphi^2\right)
\end{equation}
with three free functions, $N$, $a$ and $b$, depending on time. Such a line
element can be used to describe the spatially homogeneous Schwarzschild
interior.  Because the line element is also spherically symmetric, it is
of the general form
\begin{equation} \label{SphSymmMetric} 
{\rm d}s^2= -N(t,x)^2{\rm d}t^2+
  L(t,x)^2 \left({\rm d}x+M(t,x){\rm d}t\right)^2
+ S(t,x)^2 \left({\rm
      d}\vartheta^2+\sin^2\vartheta{\rm d}\varphi^2\right)\,,
\end{equation}
specialized to $r$-independent coefficients as well as vanishing shift, $M=0$.

Since we will use spherically symmetric models later on, we quote the
dynamical equations implied for the coefficients of (\ref{SphSymmMetric}) by a
local, generally covariant theory in which a line element of this form would
indeed correctly describe the symmetries of solutions. In the
$1+1$-dimensional context in which spherically symmetric models are placed, it
is well-known that this set of theories is given by dilaton-gravity models
\cite{Strobl,DilatonRev,NewDilaton} in which, up to field redefinitions, only
a specific set of functions, including the dilaton potential $V(S)$, can be
varied while all other terms in an action or Hamiltonian constraint are fully
determined by covariance. The equivalence of the generalized dilaton models
introduced in \cite{NewDilaton} with 2-dimensional Horndeski theories
\cite{Horndeski}, and therefore with the most general 2-dimensional local
scalar-tensor theory with second-order field equations, has recently been
demonstrated in \cite{DilatonHorndeski}. This general class of theories also
includes Palatini-$f(R)$ models \cite{Buchdahl} through their equivalence with
scalar-tensor theories with a non-dynamical scalar field \cite{PalatinifR}.

The action of any such theory can be written as 
\begin{equation}\label{S}
 {\rm S}[g,\phi]= \frac{1}{16\pi G} \int {\rm d}^2x \sqrt{-\det g}
 \bigl(\xi(\phi) R+ k(\phi,X)
+ C(\phi,X) \nabla^a\phi\nabla^b\phi
   \nabla_a\nabla_b\phi\bigr)
\end{equation}
with three free functions, $\xi(\phi)$, $k(\phi,X)$ and $C(\phi,X)$ of the
scalar field $\phi$ and
\begin{equation}
 X=-\frac{1}{2} g^{ab} \nabla_a\phi\nabla_b\phi\,.
\end{equation}
As a parameterization of the most general second-order theories in two
dimensions, the three functions $\xi(\phi)$, $k(\phi,X)$ and $C(\phi,X)$ are
not independent if field redefinitions of $\phi$ and $g_{ab}$ are allowed. For
instance, $\xi(\phi)$ can be mapped to one by a suitable $\phi$-dependent
conformal transformation of $g_{ab}$, adjusting also $k(\phi,X)$. This
ambiguity will not concern us here. It is only important that for any local
generally covariant theory for a 2-dimensional metric and a scalar field with
second-order field equations there is a choice of $\xi(\phi)$, $k(\phi,X)$ and
$C(\phi,X)$ such that the action is of the form (\ref{S}).

The canonical formulation of (\ref{S}) in this general form (and without
fixing the gauge) is rather involved because it requires inversions of some of
the free functions or their derivatives. Since the models under consideration
here are canonical, we will therefore begin with a restricted class of
spherically symmetric theories in which, compared with (\ref{S}), we have
$\xi=1$, $C=0$ and $k$ linear in $X$. That is, we will first consider
minimally coupled scalar-tensor theories with quadratic kinetic terms. In a
second step, we will then show that our result does not depend on field
redefinitions that change $\xi(\phi)$, or on an introduction of non-trivial
$k$ and $C$.

The most general covariant theory under these conditions can be derived
directly at the canonical level; see \cite{Action,MidiClass,Foundations} for
explicit derivations.  This dynamics tells us that, up to canonical
transformations, the momenta canonically conjugate to $S$ and $L$,
respectively, are given by
\begin{equation} \label{SphSymmMom}
 p_S= -\frac{1}{N} \left(\frac{\partial(SL)}{\partial t}-
   \frac{\partial(MSL)}{\partial x}\right) \quad,\quad
 p_L=
 -\frac{S}{N}\left(\frac{\partial S}{\partial t}- M \frac{\partial S}{\partial
     x}\right)\,.
\end{equation} 
The Hamiltonian constraint
\begin{equation}\label{SphSymmHam}
H_{\rm sph}[N]=\int{\rm d}xN\left(-\frac{p_Sp_L}{S}+\frac{Lp_L^2}{2S^2}
+\frac{S}{L} \frac{\partial^2 S}{\partial x^2}-
 \frac{S}{L^2}\frac{\partial S}{\partial x}\frac{\partial L}{\partial x}+ 
\frac{1}{2L}\left(\frac{\partial S}{\partial x}\right)^2+
\frac{1}{4} LSV(S)\right)
\end{equation}
and diffeomorphism constraint
\begin{equation} \label{SphSymmDiff}  
D_{\rm sph}[M]=\int{\rm d}xM\left(p_S\frac{\partial S}{\partial x}-
  L\frac{\partial 
    p_L}{\partial x}\right)
\end{equation}
then generate the equations of motion
 \begin{eqnarray} \label{pSdot}
\frac{1}{N}\frac{\partial p_S}{\partial t} &=& -\frac{p_Sp_L}{S^2}+
\frac{Lp_L^2}{S^3} 
-\frac{1}{L}\frac{\partial^2S}{\partial x^2} - \frac{1}{LN} \frac{\partial
  N}{\partial x} \frac{\partial S}{\partial x}+ \frac{1}{L^2N} \frac{\partial
  L}{\partial x} \frac{\partial(NS)}{\partial x}- \frac{S}{LN}
\frac{\partial^2N}{\partial x^2}\nonumber\\
&& -\frac{L}{4} \frac{{\rm d}(SV(S))}{{\rm d}S}
+ \frac{\partial(Mp_S)}{\partial x}
\end{eqnarray}
and
\begin{equation} \label{pLdot}
 \frac{1}{N}\frac{\partial p_L}{\partial t}= -\frac{p_L^2}{2S^2}
-\frac{S}{NL^2} \frac{\partial S}{\partial x} \frac{\partial N}{\partial
  x}- \frac{1}{2L^2} \left(\frac{\partial S}{\partial x}\right)^2-
\frac{1}{4}SV(S)+ M\frac{\partial p_L}{\partial x} 
\end{equation}
while equations for $\partial S/\partial t$ and $\partial L/\partial t$ follow
from the equations (\ref{SphSymmMom}) for the momenta.  For spherically
symmetric general relativity, the dilaton potential is given by $V(S)=-2/S$.

Since Kantowski--Sachs models are spherically symmetric, we can derive the
momenta, constraints, and equations of motion of (\ref{dsKS}) by specializing
the equations of spherical symmetry, also using $M=0$ and $x$-independence.
We obtain the momenta
\begin{equation}
 p_a=
 -\frac{b}{N}\frac{\partial b}{\partial t}
\quad,\quad p_b= -\frac{1}{N} \frac{\partial(ab)}{\partial t}\,,
\end{equation}
with Hamiltonian constraint
\begin{equation} \label{HNab}
 H_{\rm hom}[N]= N\left(-\frac{p_ap_b}{b}+ \frac{ap_a^2}{2b^2}
 - \frac{a}{2}\right)\,.
\end{equation}
It implies the equations of motion
\begin{eqnarray}
 \frac{{\rm d}a}{{\rm d}t} &=& \frac{\partial H_{\rm hom}[N]}{\partial p_a}=
 N\left(-\frac{p_b}{b}+ \frac{ap_a}{b^2}\right) \label{adot}\\
 \frac{{\rm d}b}{{\rm d}t} &=& \frac{\partial H_{\rm hom}[N]}{\partial p_b}=
 -N\frac{p_a}{b}\label{bdot}\\
 \frac{{\rm d}p_a}{{\rm d}t} &=& -\frac{\partial H_{\rm hom}[N]}{\partial a}=
 \frac{1}{2} N\left(1-\frac{p_a^2}{b^2}\right)\\
 \frac{{\rm d}p_b}{{\rm d}t} &=& -\frac{\partial H_{\rm hom}[N]}{\partial b}=
 -N\left(\frac{p_ap_b}{b^2}- \frac{ap_a^2}{b^3}\right)\,.
\end{eqnarray}

\subsection{Timelike homogeneity}
\label{s:Timelike}

In the Schwarzschild exterior, $r>R_{\rm S}$, slices of constant $r$ are still
homogeneous but timelike. The resulting canonical relationships can be derived
from the Kantowski--Sachs equations by a complex canonical transformation from
$a$, $p_a$ and $N$ to
\begin{equation} \label{ia}
 A=ia\quad,\quad p_A=-ip_a \quad,\quad n=iN
\end{equation}
while $b$ and $p_b$ remain unchanged. 
The line element (\ref{dsKS}) then takes the form
\begin{equation}\label{KSexterior}
 {\rm d}s^2= n(t)^2{\rm d}t^2- A(t)^2{\rm d}x^2+ b(t)^2
 \left({\rm  
     d}\vartheta^2+ \sin^2\vartheta{\rm d}\varphi^2\right)
\end{equation}
in which slices of constant $t$ are timelike. This line element represents the
symmetry of the exterior Schwarzschild solution, where $x$ would be the
Schwarzschild {\em time} coordinate and $t$ the Schwarzschild {\em radial}
coordinate. 

We will derive the dynamics of (\ref{KSexterior}) in a generalized form that
takes into account possible modifications from loop quantum cosmology, applied
to this homogeneous model.  These modifications are subject to a large number
of quantization ambiguities. Our result, however, will be insensitive to
ambiguities because it holds for any Hamiltonian constraint
\begin{equation}  \label{HNAb}
 H_{\rm timelike}[n]= n\left(-\frac{p_Ap_b}{b}+ \frac{Ap_A^2}{2b^2}
 + \frac{A}{2} +\delta h(A,b,p_A,p_b)\right)
\end{equation}
with a non-linear function $h(A,b,p_A,p_b)$ of the canonical variables,
multiplied by a parameter $\delta$ that vanishes in the classical limit. (In
an explicit version, both $\delta$ and $h$ would be obtained from so-called
holonomy modifications of loop quantum cosmology, which always imply a
non-linear and even non-polynomial $h$.) For $\delta=0$, the classical terms
in (\ref{HNAb}) are derived by applying the complex canonical transformation
(\ref{ia}) to (\ref{HNab}).

While $A$ and $b$ are still defined geometrically by their appearance in the
line element (\ref{KSexterior}), the new term $\delta h$ in (\ref{HNAb})
modifies the relationship between momenta and time derivatives of $A$ and $b$.
The previous equations (\ref{adot})
and (\ref{bdot}) are replaced by
\begin{eqnarray}
 \frac{{\rm d}A}{{\rm d}t} &=& \frac{\partial H_{\rm timelike}[n]}{\partial p_A}=
 n\left(-\frac{p_b}{b}+ \frac{Ap_A}{b^2}+\delta \frac{\partial
     h}{\partial p_A} \right)\\
 \frac{{\rm d}b}{{\rm d}t} &=& \frac{\partial H_{\rm timelike}[n]}{\partial p_b}=
 n\left(-\frac{p_A}{b}+ \delta \frac{\partial h}{\partial p_b}\right)\,.
\end{eqnarray}
Deriving the momenta requires an inversion of these equations, which is now
non-trivial unless $h$ is a low-order polynomial in $p_A$ and $p_b$.  For our
purposes, however, it is sufficient to invert these equations perturbatively
in $\delta$. Since our aim is to show that no modifications of this form are
compatible with slicing independence, and since an effective theory
parameterized by some $\delta$ is covariant if and only if it is covariant
order by order in $\delta$, a perturbative treatment to leading order in
$\delta$ suffices to show that the theory violates covariance. To first order
in $\delta$, we then have the simple inversion
\begin{eqnarray}
 p_A &=& -\frac{b}{n}\frac{{\rm d}b}{{\rm
       d}t}+ \delta b \frac{\partial h}{\partial p_b} \label{pia}\\
 p_b &=& -\frac{1}{n}\left(b\frac{{\rm d}A}{{\rm d}t}+A\frac{{\rm d}b}{{\rm
       d}t}\right) + \delta\left(A\frac{\partial h}{\partial p_b}+
   b\frac{\partial h}{\partial p_A}\right)\label{pib}
\end{eqnarray}
of the previous equations. We have not explicitly replaced the appearance of
$p_A$ and $p_b$ in $\delta$-terms on the right-hand sides. To first order in
$\delta$, these appearances merely represent the classical form of the momenta.

In terms of time derivatives, the modified Hamiltonian therefore equals
\begin{equation}  \label{HN2}
 H_{\rm timelike}[n]= -n \left( \frac{b}{n^2} \frac{{\rm d}A}{{\rm
       d}t}\frac{{\rm d}b}{{\rm d}t} + \frac{1}{2}A \left(\frac{1}{N^2}
     \left(\frac{{\rm d}b}{{\rm d}t}\right)^2-1\right)\right)
 +n\delta \left(- p_A\frac{\partial h}{\partial p_A}-
  p_b\frac{\partial h}{\partial 
  p_b}+h\right)\,.
\end{equation}
It is modified by a $\delta$-term if and only if $h$ is non-linear in momenta
which, to repeat, is always the case in models of loop quantum gravity. Our
main result will depend only on this general feature.

\subsection{Testing slicing independence}

If space-time is of classical Riemannian form, the condition of homogeneity in
a timelike direction is equivalent to the existence of a static solution in a
spacelike slicing. The direction in which the timelike slicing ``evolves''
then corresponds to a direction of inhomogeneity in the spacelike slicing. In
the present context, both slicings share rotational symmetry implied by the
angle-dependent spherical line element. For this statement, we do not need
complete slices, our derivations will therefore apply to local properties of
space-time. They are insensitive to any renormalization procedures that may
have to be applied to parameters in $h$ or $\delta$ in an effective theory if
fields are evolved over a wide range of scales.

The modification of (\ref{HN2}) by $\delta$-terms does not change the
symmetric nature in the geometrical interpretation of the solution as the
dynamics of a timelike slicing of space-time. If it belongs to a generally
covariant, slicing-independent theory, it must therefore allow an equivalent
description as a static solution in a spherically symmetric spacelike
slicing. The Hamiltonian (\ref{HN2}) is based on a canonical formulation with
the same phase space as the classical theory; therefore, it is a homogeneous
model of a local gravitational theory in metric variables. Non-locality would
imply additional degrees of freedom through auxiliary fields that describe
non-local terms, or higher time derivatives in a derivative expansion, but no
such fields are implied by holonomy modifications in homogeneous
models. Therefore, the corresponding spherically symmetric theory should be
local if covariance is realized. Here, it is important that we are not just
looking for an embedding of a single solution or a class of solutions in a
covariant theory, but rather have to make sure that the complete canonical
description, including the phase-space structure, can be realized in a
generally covariant theory. Similarly, the number of phase-space degrees of
freedom in holonomy-modified homogeneous models, with a single momentum per
spatial metric or triad component, implies that we are looking for a theory
with second-order field equations. No higher-derivative terms are therefore
allowed, even if they are local.

Since all local, generally covariant theories with one inhomogeneous spatial
dimension and second-order field equations are, up to field redefinitions,
given by $1+1$-dimensional dilaton-gravity models of the form (\ref{S}), where
the scalar $\phi$ is now given by the function $S$ which does transform like a
scalar under 2-dimensional coordinate transformations of $t$ and $x$, there
must be functions $\xi(S)$, $k(S,X)$ and $C(S,X)$ such that all solutions of
the dynamics generated by (\ref{HN2}) can be mapped to solutions of this
generalized dilaton-gravity theory. This condition allows us to test whether
models of loop quantum cosmology with non-linear $\delta$-terms in (\ref{HN2})
can be consistent with slicing independence in a generally covariant
theory. As already mentioned, we will first evaluate this condition in the
restricted setting in which a single dilaton potential $V(S)$ in
(\ref{SphSymmHam}) characterizes a given model.

In order to determine a possible mapping that could relate the two slicings,
we compare 
the homogeneous line element (\ref{KSexterior}) on
timelike slices with a static spherically symmetric one,
\begin{equation} \label{dsSphSymm2}
 {\rm d}s^2= -K(X)^2{\rm d}T^2+ L(X)^2 {\rm d}X^2+ S(X)^2
 \left({\rm   d}\vartheta^2+\sin^2\vartheta{\rm d}\varphi^2\right)\,.
\end{equation}
Staticity restricts the dependence of the metric components to $X$, while it
implies zero shift vector. This comparison uniquely determines the candidate
mapping 
\begin{equation}
 X=t \quad, \quad T=x
\end{equation}
for coordinates, combined with
\begin{equation} \label{abN}
 A=K\quad,\quad b=S\quad,\quad n=L
\end{equation}
for metric components.

Solutions in the homogeneous slicing must be such that the Hamiltonian
constraint $H_{\rm timelike}[n]=0$ is satisfied. In a covariant theory, the
same equation must hold true after applying the mapping (\ref{abN}), but it
need not correspond to the spherically symmetric Hamiltonian constraint. (In
fact, it does not, as we will see soon.) For covariance, it would be
sufficient if it were a combination of all the equations available in the
spherically symmetric slicing, including the staticity condition in addition
to the general spherically symmetric constraints and equations of motion.

An explicit transformation of $H_{\rm timelike}[n]=0$ to a spherically
symmetric model, using (\ref{abN}) together with a substitution of
$t$-derivatives by $X$-derivatives, shows which spherically symmetric
equations should be referred to. Transforming $H_{\rm timelike}[n]$, we obtain
the expression
\begin{equation} \label{Hhomsph}
 H_{\rm timelike}[L]=-\frac{S}{L} \frac{{\rm d}K}{{\rm
       d}X}\frac{{\rm d}S}{{\rm d}X} - \frac{1}{2}KL \left(\frac{1}{L^2}
     \left(\frac{{\rm d}S}{{\rm d}X}\right)^2-1\right)
 +L\delta \left(- p_A\frac{\partial h}{\partial p_A}-
  p_b\frac{\partial h}{\partial 
  p_b}+h\right)
\end{equation}
where
\begin{equation}
 p_A = -\frac{S}{L}\frac{{\rm d}S}{{\rm
       d}X}+ O(\delta)\quad,\quad
 p_b = -\frac{1}{L}\left(S\frac{{\rm d}K}{{\rm d}X}+K\frac{{\rm d}S}{{\rm
       d}X}\right) + O(\delta) \label{pibSL}
\end{equation}
are implied by (\ref{pia}) and (\ref{pib}). (We do not need to replace these
expressions explicitly in the $\delta$-term of (\ref{Hhomsph}), allowing us to
work with a more compact constraint.)

In the form (\ref{Hhomsph}), the constraint of the timelike slicing clearly
cannot directly correspond to the spherically symmetric Hamiltonian constraint
because it depends on the lapse function $K$ of the spherically symmetric
slicing not just through $K$ itself but also through its derivative, ${\rm
  d}K/{\rm d}X$. An additional condition is therefore required if $H_{\rm
  timelike}[L]$ is to vanish for all solutions in a {\em static} spherically
symmetric model. It turns out that staticity can be used to eliminate the
derivative ${\rm d}K/{\rm d}X$ in favor of $K$ itself and the remaining
fields, $L$ and $S$ and their derivatives. In particular, evaluating
(\ref{pLdot}) with $p_L=0$ and $M=0$, implied by staticity, (as well as $N=K$)
leads to the differential equation
\begin{equation} \label{Static}
0= -\frac{S}{KL^2} \frac{{\rm d} S}{{\rm d} X} \frac{{\rm d} K}{{\rm d}
  X}- \frac{1}{2L^2} \left(\frac{{\rm d} S}{{\rm d} X}\right)^2-
\frac{1}{4}SV(S)\,.
\end{equation}
(We need only one further condition, and therefore will not use a second
independent equation (\ref{pSdot}) in the present context. This equation is
more complicated but would give equivalent results.)  Solving this equation
{\em algebraically} for ${\rm d}K/{\rm d}X$, we can therefore eliminate this
derivative from (\ref{Hhomsph}), such that
\begin{equation} \label{HLV}
 H_{\rm timelike}[L]= \frac{1}{2} KL \left(1+ \frac{1}{2}SV(S)\right) +
 O(\delta)\,. 
\end{equation}

Disregarding $\delta$-terms, this expression vanishes in a spherically
symmetric model provided the dilaton potential indeed belongs to classical
spherically symmetric gravity, $V(S)=-2/S$. Classically, therefore, the
Hamiltonian constraint equation in the timelike homogeneous slicing amounts to
a combination of the staticity condition and a condition on the dilaton
potential in the spherically symmetric slicing. It is independent of the
spherically symmetric Hamiltonian constraint, which rather can be seen to
correspond to one of the equations of motion in the timelike homogeneous
slicing. (Recall that spatial derivatives in the spherically symmetric slicing
correspond to normal derivatives in the timelike homogeneous slicing.)

The equivalence is no longer realized if we include $\delta$-terms of the
general form as shown in (\ref{HN2}) with non-linear $h$. In particular, these
terms depend on $p_b$ which, following (\ref{pibSL}) depends on the lapse
function $K$ of a spherically symmetric slicing. If $h$ is non-linear in
$p_b$, the $\delta$-terms are non-linear in $p_b$, or non-linear in $K$ after
the transformation to spherically symmetric variables. After factoring out a
single factor of $KL$, as in (\ref{HLV}), the remaining terms $H_{\rm
  timelike}[L]/(KL)$ therefore still depend on $K$. If the dynamics in this
slicing corresponds to a dilaton model with Hamiltonian (\ref{SphSymmHam}),
$H_{\rm timelike}[L]/(KL)$ can depend only on $S$ because the dilaton
potential is restricted by the covariance condition to have only such a
dependence.

However, if $H_{\rm timelike}[L]/(KL)$ depends on $K$ when $\delta$-terms are
included, it cannot just depend on $S$: While the staticity condition
(\ref{Static}) can be used to solve for $K$ in terms of $S$ and $L$, it
requires solving the differential equation; an algebraic solution for ${\rm
  d}K/{\rm d}X$ as in the classical case is not sufficient. A solution $K$ of
(\ref{Static}) depends on $S$ and $L$ non-locally because integrations are
required. No non-local $S$-dependence, and no dependence on $L$ at all, can be
absorbed in a local dilaton potential $V(S)$. Therefore, there is no local
generally covariant theory of the restricted form considered so far, that
could describe a spherically symmetric slicing corresponding to the timelike
homogeneous one which, by construction, is also local. Any $\delta$-term with
non-linear $h$ therefore violates slicing independence and general covariance.

So far, we have shown that there is no minimally coupled generally covariant
theory quadratic in momenta which could correspond to the modified dynamics of
a timelike homogeneous slicing. It is not difficult to see that non-minimal
coupling, leading to $\xi\not=1$ in (\ref{S}), does not change the
result. Such a theory can always be obtained from a minimally coupled one by a
field redefinition, using an $S$-dependent conformal transformation of the
2-dimensional metric. Such a transformation, formulated canonically, would
rescale $K$ and $L$ by $S$-dependent functions, such that there would be new
terms in the equations of motion with spatial derivatives of $S$, but no new
derivative terms of $K$ or $L$. It is impossible for such terms to absorb a
$K$-dependence in a $\delta$-term as mentioned in the preceding paragraph, or
a term non-local in $L$ or with an entire derivative expansion of $L$ if a
solution of (\ref{Static}) for $K$ is used. Our no-go result therefore extends
to non-minimally coupled scalars. Similarly, allowing for terms with
non-linear $k$ or non-zero $C$ in (\ref{S}) leads, in the static case which is
relevant here, only to terms with additional spatial derivatives of $S$ or at
most first-order derivatives of $L$ through the Christoffel symbol required in
the $C$-term of (\ref{S}). Again, even with the freedom of choosing $k$ or $C$
it is impossible to absorb the dependence on $K$ or non-locally on $L$ that
results from a $\delta$-term.

Our no-go theorem is therefore complete: There is no generally covariant
spherically symmetric theory that could have a solution space corresponding to
the modified timelike homogeneous model. Slicing independence is therefore
violated by holonomy modifications in symmetry-reduced models of loop quantum
gravity.

\section{Conclusion}

We have shown that the model proposed in \cite{Transfig} violates general
covariance and therefore fails to describe space-time or black holes. This
result has implications even if one is not interested in black holes but only
in cosmological applications of models of loop quantum gravity. If such models
are sufficiently general, it must be possible to apply any proposed
modification to models with the symmetries of Kantowski--Sachs space-times,
including those with a timelike homogeneous slicing.  If they belong to a
generally covariant theory, it must then be possible to find a consistent
mapping to a static spherically symmetric slicing. Our results show that this
is never the case for holonomy modifications proposed in models of loop
quantum cosmology. This is our no-go result about general covariance in this
setting. (Our result is consistent with the observation that all proposed
analog actions that could describe holonomy modifications by higher-curvature
terms in isotropic models \cite{LimCurvLQC,HigherDerivLQC}, based on mimetic
gravity \cite{Mimetic,MimeticRev}, fail to describe related effects in
anisotropic models \cite{MimeticLQC} or for perturbative inhomogeneity
\cite{MimeticLQCPert}.)

Such a general violation of covariance might be interpreted as ruling out not
only a specific model but also the entire approach, based on loop
quantization. Luckily, however, previous research had already independently
shown a possible way out of this damning conclusion. It is possible to evade
our no-go result if one takes into account the possibility of {\em
  generalized} space-time structures that may be considered covariant in the
sense that the same number of gauge transformations is realized as in the
classical theory, but in a way that no longer corresponds to slicing
independence in Riemannian geometry
\cite{Action,SphSymmCov}. In fact, the constructions of
\cite{Transfig} implicitly assumed that space-time, even after modifying
dynamical equations originally derived from general relativity, retains its
Riemannian structure and can be described by line elements. Our no-go result
rules out this implicit assumption, but it may be evaded if the assumption is
relaxed.

It is difficult to describe non-Riemannian space-time structures in general
terms because most of our intuition in general relativity is built on
Riemannian properties. Nevertheless, in canonical theories, there are
systematic methods that allow one to test whether gauge symmetries are
respected by modifications or quantization, even while algebraic relations of
the symmetries may be subject to modifications themselves. General covariance
in canonical gravity is expressed as the condition of anomaly-freedom of the
constraints that generate hypersurface deformations in space-time, given by
the Hamiltonian and diffeomorphism constraints. If modified constraints still
have closed Poisson brackets, the theory is anomaly-free and enjoys the same
number of gauge transformations as classical general relativity, with
hypersurface deformations realized in the classical limit. If this is the
case, the theory may be considered covariant but in a generalized way. The
modified constraints then may no longer generate hypersurface deformations in
space-time, but they do provide a well-defined set of gauge transformations
that allow one to remove the correct number of spurious degrees of
freedom. Line elements might then be applicable in certain regions of
space-time, after a field redefinition of metric components that in certain
cases can be derived from the consistent generators of modified
hypersurface deformations \cite{Normal,EffLine}. These constructions are now
being investigated \cite{DefSchwarzschild,DefSchwarzschild2,DefGenBH}, but
generalized covariance must be better understood before it is possible to
derive complete and reliable models of black holes in loop quantum gravity.

\section*{Acknowledgements}

This work was supported in part by NSF grant PHY-1912168.


\end{document}